# Biomolecular events in cancer revealed by attractor metagenes

Wei-Yi Cheng[1] & Dimitris Anastassiou[1]

**Mining gene expression profiles has proven valuable for identifying metagenes, defined as linear combinations of individual genes, serving as surrogates of biological phenotypes. Typically, such metagenes are jointly generated as the result of an optimization process for dimensionality reduction. Here we present an unconstrained method for individually generating metagenes that can point to the core of the underlying biological mechanisms. We use an iterative process that starts from any seed gene and converges to one of several precise attractor metagenes representing biomolecular events, such as cell transdifferentiation or the presence of an amplicon. By analyzing six rich gene expression datasets from three different cancer types, we identified many such biomolecular events, some of which are present in all tested cancer types. We focus on several such events including a stage-associated mesenchymal transition and a grade-associated mitotic chromosomal instability.**

Rich biomolecular datasets, publicly available at an increasing rate from sources such as The Cancer Genome Atlas (TCGA), provide unique opportunities for biological discovery from purely computational analysis. Gene expression signatures resulting from analysis of cancer datasets can serve as surrogates of cancer phenotypes[1]. Subtypes in many cancer types[2-4] have been identified by gene expression analysis often using techniques such as nonnegative matrix factorization[5] combined with consensus clustering[6].

The main objective addressed by techniques such as nonnegative matrix factorization is to reduce dimensionality by identifying a number of metagenes jointly representing the gene expression dataset as accurately as possible, in lieu of the whole set of individual genes. Each metagene is defined as a positive linear combination of the individual genes, so that its expression level is an accordingly weighted average of the expression levels of the individual genes. The identity of each resulting metagene is influenced by the presence of other metagenes within the objective of overall dimensionality reduction achieved by joint optimization.

In contrast, if the aim is not dimensionality reduction or classification into subtypes, but instead the independent and unconstrained identification of metagenes as surrogates of pure biomolecular events, then a different algorithm should be devised.

[1] Center for Computational Biology and Bioinformatics and Department of Electrical Engineering, Columbia University, 1312 S.W. Mudd Building, Mail Code 4712, 500 West 1120th Street, New York, NY 10027, USA. Correspondence should be addressed to D. Anastassiou (anastas@ee.columbia.edu)

This approach is devoid of cross-interference or mutual exclusivity requirements and has the advantage of increasing the chance of precisely identifying the particular genes at the core of the underlying biological mechanism as those that have the highest weights in the corresponding metagene, thus shedding more light on that mechanism. We found that, given a rich dataset represented by a gene expression matrix, such surrogate metagenes can be naturally identified as stable and precise attractors using a simple iterative approach. This identification is totally unsupervised, as it does not make use of any phenotypic association. Once identified, however, a metagene attractor is likely to be found associated with a phenotype.

We found that several attractor metagenes are present in nearly identical form in multiple cancer types. This provides an additional opportunity to combine the powers of a large number of rich datasets to focus, at an even sharper level, on the core genes of the underlying mechanism. For example, this methodology can precisely point to the causal (driver) oncogenes within amplicons to be among very few candidate genes. Importantly, this can be done from rich gene expression data, which already exist in abundance, without making any use of sequencing data.

We identified many attractors, which can directly lead to corresponding testable biological hypotheses. For the purposes of this paper we present the general methodology for the benefit of the research community together with a listing of the attractors in six datasets from three cancer types (ovarian, colon, breast). We also identified other attractors in other cancer types, such as leukemia, consistent with known facts about the nature of each disease. Here, we only focus on a few interesting cancer-associated attractors that we found present in these three solid cancer types.

**RESULTS**

**Derivation of Attractor Metagenes**

Given a nonnegative measure $J(G_i, G_j)$ of pairwise association between genes $G_i$ and $G_j$, we define an attractor metagene $M = \sum_i w_i G_i$ to be a linear combination of the individual genes with weights $w_i = J(G_i, M)$. The association measure $J$ is assumed to have minimum possible value 0 and maximum possible value 1, so the same is true for the weights. It is also assumed to be scale-invariant, therefore it is not necessary for the weights to be normalized so that they add to 1, and the metagenes can still be thought of as expressing a normalized weighted average of the expression levels of the individual genes.

According to this definition, the genes with the highest weights in an attractor metagene will have the highest association with the metagene (and, by implication, they will tend to be highly associated among themselves) and so they will often represent a biomolecular event reflected by the co-expression of these top genes. This can happen, e.g., when a biological mechanism is activated, or when a copy number variation (CNV), such as an amplicon, is present, in some of the samples included in the expression matrix. In the following we use the term "attractor" for simplicity to refer to an attractor metagene, and the term "top genes" to refer to the genes with the highest weights in the attractor.

The definition of an attractor metagene can readily be generalized to include features other than gene expression, such as methylation values. It can also be used in datasets of any objects (not necessarily genes) characterized by any type of feature vectors, with applications in other disciplines, such as social and economic sciences.

The computational problem of identifying attractor metagenes given an expression matrix can be addressed heuristically using a simple iterative process: Starting from a particular seed (or "attractee") metagene $M$, a new metagene is defined in which the new weights are $w_i = J(G_i, M)$. The same process is then repeated in the next iteration resulting in a new set of weights, and so forth. In all gene expression datasets that we tried we found that this process converges to a limited number of stable attractors. Each attractor is defined by a precise set of weights, which are reached with high accuracy typically within 10 or 20 iterations.

This algorithmic behavior with nice convergence properties is not surprising, because if a metagene contains some co-expressed genes with high weights, then the next iteration will naturally "attract" even more genes with the same properties, and so forth, until the process will eventually converge to a metagene representing a potential underlying biological event reflected by this co-expression. Furthermore, the set of the few genes with the highest weight are likely to represent the "heart" (core) of the biomolecular event. In support of this concept, the association of any of the top-ranked individual genes with the attractor metagene is consistently and significantly higher than the pairwise association between any of these genes, suggesting that the set of these top genes jointly comprise a proxy representing a biomolecular event better than each of the individual genes would. The top genes in the attractors can also serve as readily derived modules and we can use them in a complementary manner within the context of methods of identifying regulatory modules or other related approaches[7].

Indeed, related versions of the signatures identified by attractors in this paper have been previously identified in various contexts, often intermingled with additional genes that may be unrelated or weakly related to the main underlying mechanism. However, the contribution of our work is that these signatures are recognized as multi-cancer biomolecular events, sharply pointing to the underlying mechanism. Therefore the top genes of the attractors will be appropriate for being used as biomarkers or for understanding the underlying biology. For example, one of the attractors that we identified (the "mitotic CIN" attractor, described below) has previously been found among sets of genes described generally[8] as "proliferation" or "cell cycle related" markers, while the actual attractor points much more sharply to particular elements in the structure of the kinetochore.

A reasonable implementation of an "exhaustive" search is to only consider the seed metagenes in which one selected "attractee" gene is assigned a weight of 1 and all the other genes are assigned a weight of 0. The metagene resulting from the next iteration will then assign high weights to all genes highly associated with the originally selected gene, to which we refer as the "attractee gene." In this way all attractors representing biomolecular events characterized by coordinately co-expressed genes will be identified when these genes are used as attractees. The computational implementation of the algorithm is described in Online Methods. We note that a dual method can be used to identify attractor "metasamples" as representatives of subtypes, and we can also combine such metasamples with the attractor metagenes in various ways to achieve biclustering, but this topic is not examined in this paper.

We analyzed six datasets, two from ovarian cancer, two from breast cancer and two from colon cancer:

| Dataset | Sample Size | Platform |
| --- | --- | --- |
| Breast Wang (GSE2034) | 286 | Affymetrix HG-U133A |
| Breast TCGA | 536 | Agilent 244K Custom Gene Expression G4502A-07-3 |
| Colon Jorrison (GSE14333) | 290 | Affymetrix HG-U133Plus 2.0 |
| Colon TCGA | 154 | Agilent 244K Custom Gene Expression G4502A-07-3 |
| Ovarian Tothill (GSE9891) | 285 | Affymetrix HG-U133Plus 2.0 |
| Ovarian TCGA | 584 | Affymetrix HG-U133A |

In each case, we identified general and genomically localized attractors and we found that many among them appear in similar forms in all six datasets. Following are descriptions of some of our results, starting with the three strongest multi-cancer attractors.

**Mesenchymal Transition Attractor**

This attractor contains mostly epithelial-mesenchymal transition (EMT)-associated genes. Table 1 provides a listing of top 100 genes based on their average mutual information (Online Methods) with their corresponding attractor metagenes.

This is a stage-associated attractor, in which the signature is significantly present only when a particular level of invasive stage, specific to each cancer type, has been reached. For example, this association can be demonstrated in three cancer datasets from different types (breast GSE3893, ovarian TCGA and colon GSE14333) that were annotated with clinical staging information: We can create a listing of differentially expressed genes, ranked by fold change, when ductal carcinoma in situ (DCIS) progresses to invasive ductal carcinoma; ovarian cancer progresses to stage III; and colon cancer progresses to stage II. In all three cases, the attractor is highly enriched among the top genes. Specifically, among the top 100 differentially expressed genes, the number of attractor genes included in Table 1 is 55 in breast cancer, 45 in ovarian cancer and 31 in colon cancer. The corresponding $P$ values are $3\times10^{-109}$, $9\times10^{-83}$ and $5\times10^{-62}$, respectively.

This attractor has been previously identified with remarkable accuracy as representing a particular kind of mesenchymal transition of cancer cells present in all types of solid cancers tested leading to a published list of top 64 genes[9, 10]. Indeed 56 of these top 64 genes also appear in Table 1 ($P < 10^{-127}$), and furthermore all top 24 genes of Table 1 are among the 64. We found that most of the genes of the signature were expressed by the cancer cells themselves, and not by the surrounding stroma, at least in a neuroblastoma xenograft model that we tried[10]. We also found that the signature is associated with prolonged time to recurrence in glioblastoma[11]. Related versions of the same signature were previously found to be associated with resistance to neoadjuvant therapy in breast cancer[12]. These results are consistent with the finding that EMT induces cancer cells to acquire stem cell properties[13]. It has been hypothesized that EMT is a key mechanism for cancer cell invasiveness and motility[14-16]. The attractor, however, appears to represent a more general phenomenon of transdifferentiation present even in nonepithelial cancers such as neuroblastoma, glioblastoma and Ewing's sarcoma.

Although similar signatures are often labeled as "stromal," because they contain many stromal markers such as α-SMA and fibroblast activation protein, the fact that most of the genes of the signature were expressed by xenografted cancer cells[10], and not by mouse stromal cells, suggests that this particular attractor of coordinately expressed genes represents cancer cells having undergone a mesenchymal transition. The signature may indicate a non-fibroblastic transition, as occurs in glioblastoma, in which case collagen *COL11A1* is not co-expressed with the other genes of the attractor. We have hypothesized that a full fibroblastic transition of the cancer cells occurs when cancer cells encounter adipocytes[10], in which case they may well assume the duties of cancer-associated fibroblasts (CAFs) in some tumors[17]. In that case, the best proxy of the signature[9] is *COL11A1* and the strongly co-expressed genes *THBS2* and *INHBA*. Indeed, the 64 genes of the previously identified signature were found from multi-cancer analysis[9] as the genes whose expression is consistently most associated with that of *COL11A1*.

The only EMT-inducing transcription factor found upregulated in the xenograft model[10] is SNAI2 (Slug), and it is also the one most associated with the signature in publicly available datasets. We also found that the microRNAs most highly associated with this attractor are miR-214, miR-199a, and miR-199b. Interestingly, miR-214 and miR-199a were found to be jointly regulated by another EMT-inducing transcription factor, TWIST1[18].

**Mitotic CIN Attractor**

This attractor contains mostly kinetochore-associated genes. Table 2 provides a listing of the top 100 genes based on their average mutual information with their corresponding attractor metagenes, starting from *CENPA*, which encodes for a histone H3-like centromeric protein.

Contrary to the stage-associated mesenchymal transition attractor, this is a grade-associated attractor, in which the signature is significantly present only when an intermediate level of tumor grade is reached. For example, this association can be demonstrated in three cancer datasets from different types (breast GSE3494, ovarian TCGA and bladder GSE13507) that were annotated with tumor grade information. We can create a listing of differentially expressed genes, ranked by fold change, when grade G2 is reached. In all three cases, the attractor is highly enriched among the top genes. Specifically, among the top 100 differentially expressed genes, the number of attractor genes included Table 2 is 41 in breast cancer, 36 in ovarian cancer and 26 in colon cancer. The corresponding $P$ values are $7 \times 10^{-73}$, $4 \times 10^{-61}$ and $5 \times 10^{-47}$, respectively. Consistently, a similar "gene expression grade index" signature[19] was previously found differentially expressed between histologic grade 3 and histologic grade 1 breast cancer samples. Furthermore, that same signature[19] was found capable of reclassifying patients with histologic grade 2 tumors into two groups with high versus low risks of recurrence.

This attractor is associated with chromosomal instability (CIN), as evidenced from the fact that another similar gene set comprising a "signature of chromosomal instability"[20] was previously derived from multiple cancer datasets purely by identifying the genes that are most correlated with a measure of aneuploidy in tumor samples. This led to a 70-gene signature referred to as "CIN70." Indeed 34 of these 70 genes appear in Table 2 ($P < 10^{-61}$). However, several top genes of the attractor, such as *CENPA*, *KIF2C*, *BUB1* and *CCNA2* are not present in the CIN70 list. Mitotic CIN is increasingly recognized[21] as a widespread multi-cancer phenomenon.

The attractor is characterized by overexpression of kinetochore-associated genes, which is known[22] to induce CIN for reasons that are not clear. Overexpression of several of the genes of the attractor, such as the top gene *CENPA*[23], as well as *MAD2L1*[24] and *TPX2*[25], has also been independently previously found associated with CIN. Included in the mitotic CIN attractor are key components of mitotic checkpoint signaling[26], such as BUB1B, MAD2L1 (aka MAD2), CDC20, and TTK (aka MSP1). Also among the genes in the attractor is *MKI67* (aka *Ki-67*), which has been widely used as a proliferation rate marker in cancer.

Among transcription factors, we found *MYBL2* (aka *B-Myb*) and *FOXM1* to be strongly associated with the attractor. They are already known to be sequentially recruited to promote late cell cycle gene expression[27] to prepare for mitosis.

Inactivation of the retinoblastoma (RB) tumor suppressor promotes CIN[28] and the expression of the attractor signature. Indeed, a similar expression of a "proliferation gene cluster[29]" was found strongly associated with the human papillomavirus E7 oncogene, which abrogates RB protein function and activates E2F-regulated genes. Consistently, many among the genes of the attractor correspond to E2F pathway genes controlling cell division or proliferation. Among the E2F transcription factors, we found that E2F8 and E2F7 are most strongly associated with the attractor.

**A lymphocyte-specific attractor**

This attractor consists mainly of lymphocyte-specific genes with prominent presence of *CD53*, *PTPRC*, *LAPTM5*, *DOCK2* and *LCP2*. It is strongly associated[30] with the expression of microRNA miR-142 as well as with particular hypermethylated and hypomethylated gene signatures, which we reconfirmed to be consistent methylation attractors in cancer methylation data with our methodology. There is also significant overlap between the sets of hypomethylated and overexpressed genes, suggesting that their expression is triggered by hypomethylation. Gene set enrichment analysis reveals that the attractor is found enriched in genes known to be preferentially expressed in differentiation into lymphocytes[31] and is also found occasionally upregulated in various cancers. Table 3 provides a listing of the top 100 genes of the attractor based on their average mutual information with their corresponding attractor metagenes.

**Chr8q24.3 amplicon attractor**

Amplification in chr8q24 is often associated with cancer because of the presence of the *MYC* (aka *c-Myc*) oncogene at location 8q24.21. Indeed, *MYC* is one of 157 genes in "amplicon 8q23-q24" previously identified[32] in an extensive study of the breast cancer "amplicome" derived from 191 samples.

We found, however, that the core of the amplified genes occurs at location 8q24.3 and this is, in fact, our most prominent multi-cancer amplicon attractor. The main core gene of the attractor appears to be *PUF60* (aka *FIR*). Other consistently present top genes are *EXOSC4*, *CYC1*, *SHARPIN*, *HSF1*, *GPR172A*. It is known that PUF60 can repress c-Myc via its far upstream element (FUSE), although a particular isoform was found to have the opposite effect[33]. The other genes may also play important roles. For example, HSF1 (heat shock transcription factor 1) has been associated with cancer in various ways[34]. It was found[35] that HSF1 can induce genomic instability through direct interaction with CDC20, a key gene of the mitotic CIN attractor mentioned above (listed in Table 2). Furthermore, HSF1 was found[36] required for the cell transformation and tumorigenesis induced by the *ERBB2* (aka *HER2*) oncogene (see subsequent discussion of *HER2* amplicon) responsible for aggressive breast tumors.

The top ten genes of the chr8q24.3 attractor, ranked by the average of the highest five values of mutual information, are shown in Table 4. Interestingly, one of our identified general attractor corresponds to an aneuploidy involving a whole arm amplification of chr8q, which is occasionally present in multiple cancer types, and this 8q amplification is the most prominent such aneuploidy attractor among all chromosomes.

**Chr17q12 HER2 amplicon attractor**

This amplicon is prominent in breast cancer[37] and we also found it present in some samples of ovarian cancer, but not as much in colon cancer. So we initially used the four data sets of breast and ovarian cancer for deriving the attractor. We found that *ERBB2* (aka *HER2*), *STAR3*, *GRB7* and *PGAP3* were the top-ranked genes, consistent with their known presence in the amplicon. We also found that gene *MIEN1* (aka *C17orf37*) was very highly ranked in the two data sets in which its probe set was present. *MIEN1* has recently been identified as an important gene within the 17q12 amplicon in various cancers including prostate cancer[38]. Therefore, we augmented the choice of datasets to the following seven, of which *MIEN1* is included in five: breast GSE2034, breast GSE32646, breast GSE36771, breast TCGA, ovarian GSE9891, ovarian GSE26193, ovarian TCGA. Table 4 shows the top ten genes ranked by the average of the top five scores of mutual information in the seven datasets for each gene. The results suggest that the above-mentioned five genes, including *MIEN1*, are consistently strongly co-expressed, and therefore are likely "driver" genes in the amplicon.

In addition to the narrow *HER2* amplicon, it is known that sometimes a large amplicon extends to more than a million bases containing both *HER2* as well as *TOP2A* (one of the genes of the mitotic CIN attractor) at 17q21[39]. We have observed that *TOP2A* indeed appears among the top 50 genes in terms of its association with the attractor in breast cancer. *HER2*/*TOP2A* co-amplification has been linked with better clinical response to therapy.

**Estrogen receptor breast cancer attractor**

We found this attractor clearly present only in breast cancer, and therefore we derived it using six breast cancer data sets (GSE2034, GSE3494, GSE31448, GSE32646, GSE36771, breast TCGA). Table 5 shows the top 50 genes ranked by the average mutual information in these datasets, revealing that genes *CA12*, *AGR2*, *GATA3*, *FOXA1*, *MLPH* and *TBC1D9* are strongly co-expressed with the estrogen receptor *ESR1* in the attractor.

**Using attractor metagenes as proxies of biomolecular events**

A biomolecular event, whether it is present in multiple cancer types or it is cancer specific, can be represented by a "consensus attractor metagene" after analyzing multiple datasets. To generate such consensus attractors, we rank individual genes in terms of their average mutual information (Online Methods) with the corresponding attractor metagenes across all datasets.

For example, Figure 1 clarifies the previously reported[40] observation that breast tumors with high chromosomal instability are predominantly of the estrogen receptor negative phenotype. The scatter plots of the mitotic CIN and estrogen receptor metagenes shown in Figure 1, however, demonstrate that the nature of the association is that ER negative tumors have high mitotic chromosomal instability (or equivalently that low chromosomal instability implies that the tumor is ER positive), but the reverse relationship is not as clear.

**DISCUSSION**

Gene expression analysis has resulted in several cancer types being further classified into subtypes labeled, e.g. as "mesenchymal" or "proliferative." Such characterizations, however, may sometimes simply reflect the presence of the mesenchymal transition attractor or the mitotic chromosomal instability attractor, respectively, in some of the analyzed samples. Similar subtype characterizations across cancer types often share several common genes, but the consistency of these similarities has not been significantly high.

In contrast, using an unconstrained algorithm independent of subtype classification or dimensionality reduction, we identified several attractors exhibiting remarkable consistency across many cancer types, suggesting that each of them represents a precise biological phenomenon present in multiple cancers.

We found that the mesenchymal transition attractor is significantly present only in samples whose stage designation has exceeded a threshold, but not in all of such samples. Similarly, we found that the mitotic chromosomal instability attractor is significantly present only in samples whose grade designation has exceeded a threshold, but not in all of them. On the other hand, the absence of the mesenchymal transition attractor in a profiled high-stage sample (or the absence of the mitotic chromosomal instability attractor in a profiled high-grade sample) does not necessarily mean that the attractor is not present in other locations of the same tumor. Indeed, it is increasingly appreciated[41] that tumors are highly heterogeneous. Therefore it is possible for the same tumor to contain components, in which, e.g., some are migratory having undergone mesenchymal transition, some other ones are highly proliferative, etc. If so, attempts for subtype classification based on one particular site in a sample may be confusing.

Similarly, existing molecular marker products make use of multigene assays that have been derived from phenotypic associations in particular cancer types. For breast cancer, biomarkers such as Oncotype DX[42] and Mammaprint[43] contain several genes highly ranked in our attractors. For example, most of the genes used for the Oncotype DX breast cancer recurrence score directly converge to one of our identified attractors: *MMP11* to the mesenchymal transition attractor; *MKI67* (aka *Ki-67*), *AURKA* (aka *STK15*), *BIRC5* (aka *Survivin*), *CCNB1*, and *MYBL2* to the mitotic CIN attractor; *CD68* to the lymphocyte-specific attractor; *ERBB2* and *GRB7* to the HER2 amplicon attractor; and *ESR1*, *SCUBE2*, *PGR* to the estrogen receptor attractor.

We envision, instead, a "multidimensional" biomarker product that will be applicable to multiple cancer types. Each of the dimensions will correspond to a specific attractor detected from a sharp choice of the genes at its core, reflecting a precise biological attribute of cancer. For example, each relevant amplicon can be identified by the coordinate co-expression of the top few genes of the attractor without any need for sequencing, and each amplicon will correspond to another dimension. The collection of the independent results in many dimensions will provide a clearer diagnostic and prognostic image after cleanly distinguishing the contributions of each component. Even though molecular marker genes in existing products are often separated into groups that are related to our attractor designation, any improvement in diagnostic, prognostic, or predictive accuracy resulting from better such group designation and better choice of genes in each group would be highly desirable. We hope that the computational methodology of identifying the attractors of cancer, as presented here, will be valuable in that regard.

# Figures

**Figure 1: Scatter plots demonstrating the relationship between mitotic CIN attractor and estrogen receptor attractor in breast cancer**. The two metagenes were defined to be "consensus attractors" after ranking individual genes in terms of their average mutual information with the corresponding attractor metagenes, across all datasets, and selecting the genes having average mutual information greater than 0.5. These criteria led to 47 genes in the consensus mitotic CIN attractor (the top 47 genes in Table 2), and *ESR1*, *CA12*, *AGR2*, *GATA3*, *FOXA1*, *MLPH* and *TBC1D9* (the top seven genes in Table 5) in the consensus estrogen receptor breast cancer attractor. These scatter plots reveal that ER-negative breast tumors have high mitotic chromosomal instability, but not necessarily vice versa.

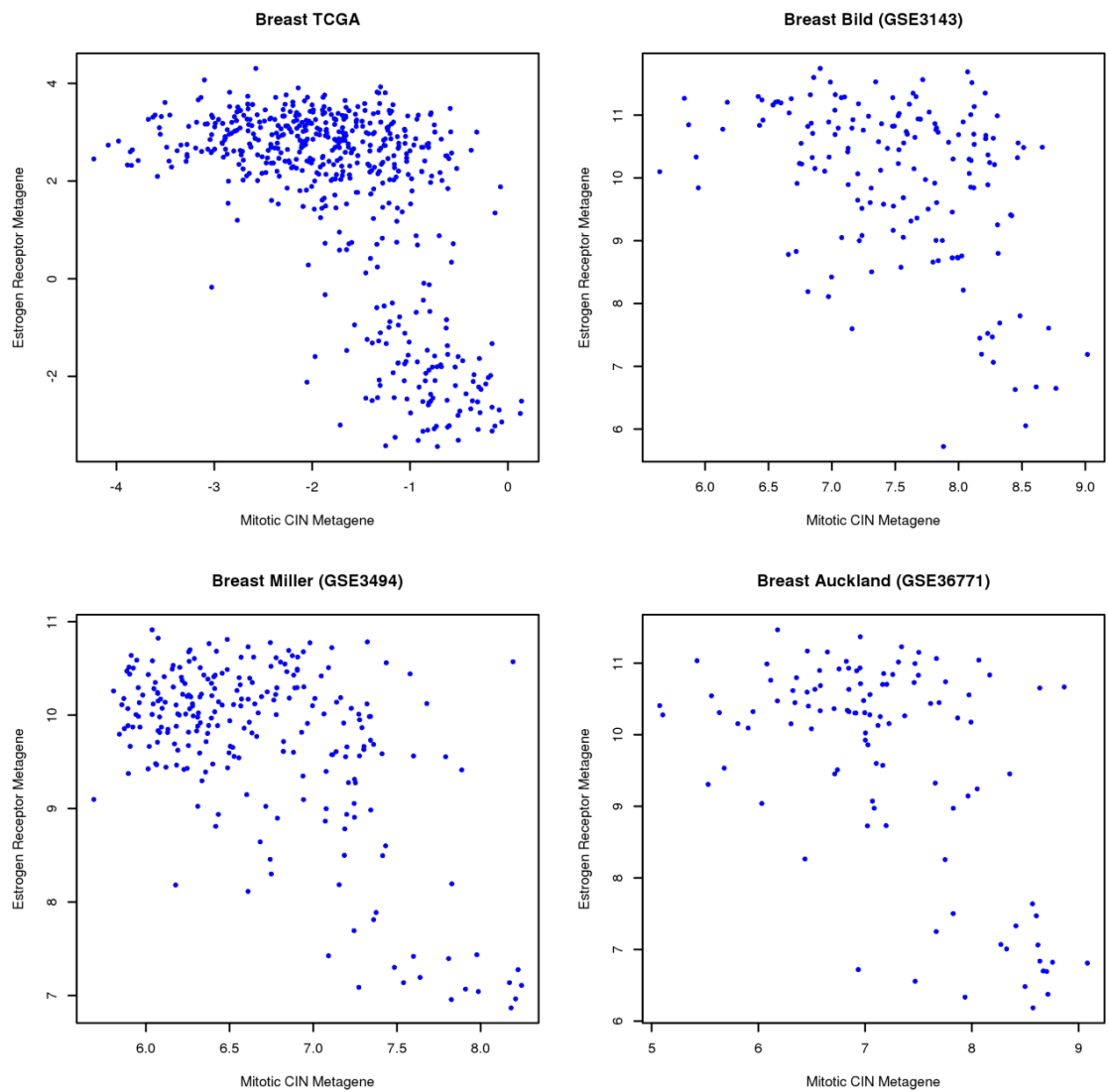

# Tables

**Table 1**: Top 100 genes of the mesenchymal transition attractor based on six datasets

| Rank | Gene | Avg MI | Rank | Gene | Avg MI | Rank | Gene | Avg MI |
|---|---|---|---|---|---|---|---|---|
| 1 | COL5A2 | 0.814 | 34 | FN1 | 0.545 | 67 | SPON2 | 0.444 |
| 2 | VCAN | 0.776 | 35 | LRRC15 | 0.534 | 68 | SPOCK1 | 0.444 |
| 3 | SPARC | 0.767 | 36 | COL11A1 | 0.529 | 69 | COL8A2 | 0.441 |
| 4 | THBS2 | 0.759 | 37 | ANTXR1 | 0.529 | 70 | PDPN | 0.437 |
| 5 | FBN1 | 0.750 | 38 | RAB31 | 0.527 | 71 | LHFP | 0.436 |
| 6 | COL1A2 | 0.750 | 39 | THY1 | 0.519 | 72 | GREM1 | 0.436 |
| 7 | COL5A1 | 0.747 | 40 | NNMT | 0.518 | 73 | GFPT2 | 0.436 |
| 8 | FAP | 0.734 | 41 | SULF1 | 0.506 | 74 | TGFB1I1 | 0.435 |
| 9 | AEBP1 | 0.712 | 42 | LOXL1 | 0.502 | 75 | C1S | 0.433 |
| 10 | CTSK | 0.709 | 43 | PRRX1 | 0.502 | 76 | EDNRA | 0.433 |
| 11 | COL3A1 | 0.688 | 44 | COL10A1 | 0.499 | 77 | GAS1 | 0.431 |
| 12 | COL1A1 | 0.684 | 45 | MXRA8 | 0.494 | 78 | NOX4 | 0.431 |
| 13 | SERPINF1 | 0.674 | 46 | WISP1 | 0.484 | 79 | FBLN2 | 0.429 |
| 14 | COL6A3 | 0.670 | 47 | RCN3 | 0.483 | 80 | TCF4 | 0.428 |
| 15 | CDH11 | 0.663 | 48 | TNFAIP6 | 0.481 | 81 | NUAK1 | 0.428 |
| 16 | GLT8D2 | 0.658 | 49 | ECM2 | 0.480 | 82 | ADAMTS2 | 0.422 |
| 17 | MMP2 | 0.654 | 50 | HTRA1 | 0.480 | 83 | NDN | 0.419 |
| 18 | LUM | 0.654 | 51 | EFEMP2 | 0.478 | 84 | DPYSL3 | 0.418 |
| 19 | DCN | 0.650 | 52 | MXRA5 | 0.474 | 85 | LAMA4 | 0.418 |
| 20 | POSTN | 0.631 | 53 | ACTA2 | 0.472 | 86 | MFAP5 | 0.418 |
| 21 | ADAM12 | 0.614 | 54 | LOX | 0.470 | 87 | LGALS1 | 0.417 |
| 22 | COL6A2 | 0.609 | 55 | ITGBL1 | 0.466 | 88 | PLAU | 0.417 |
| 23 | OLFML2B | 0.607 | 56 | PMP22 | 0.465 | 89 | DSE | 0.417 |
| 24 | INHBA | 0.601 | 57 | PTRF | 0.463 | 90 | COLEC12 | 0.416 |
| 25 | FSTL1 | 0.600 | 58 | CALD1 | 0.460 | 91 | DACT1 | 0.415 |
| 26 | SNAI2 | 0.577 | 59 | HEG1 | 0.458 | 92 | CD248 | 0.402 |
| 27 | CRISPLD2 | 0.574 | 60 | NID2 | 0.455 | 93 | COL6A1 | 0.402 |
| 28 | PDGFRB | 0.567 | 61 | TAGLN | 0.455 | 94 | RARRES2 | 0.401 |
| 29 | PCOLCE | 0.566 | 62 | SFRP4 | 0.451 | 95 | JAM3 | 0.398 |
| 30 | BGN | 0.566 | 63 | PALLD | 0.450 | 96 | SGCD | 0.397 |
| 31 | ANGPTL2 | 0.555 | 64 | OLFML1 | 0.448 | 97 | EMILIN1 | 0.394 |
| 32 | COPZ2 | 0.553 | 65 | FILIP1L | 0.447 | 98 | ZFPM2 | 0.394 |
| 33 | ASPN | 0.547 | 66 | TIMP3 | 0.445 | 99 | IGFBP7 | 0.394 |
|  |  |  |  |  |  | 100 | ZEB1 | 0.394 |

**Table 2**: Top 100 genes of the mitotic CIN attractor based on six datasets

| Rank | Gene | Avg MI | Rank | Gene | Avg MI | Rank | Gene | Avg MI |
|---|---|---|---|---|---|---|---|---|
| 1 | CENPA | 0.727 | 34 | EXO1 | 0.539 | 67 | ESPL1 | 0.448 |
| 2 | MELK | 0.679 | 35 | AURKA | 0.538 | 68 | FAM64A | 0.442 |
| 3 | KIF2C | 0.667 | 36 | CDKN3 | 0.538 | 69 | SPAG5 | 0.436 |
| 4 | BUB1 | 0.664 | 37 | DEPDC1 | 0.536 | 70 | MYBL2 | 0.436 |
| 5 | KIF4A | 0.660 | 38 | RRM2 | 0.535 | 71 | EZH2 | 0.432 |
| 6 | CCNB2 | 0.659 | 39 | CDCA8 | 0.535 | 72 | SMC4 | 0.432 |
| 7 | KIF20A | 0.657 | 40 | SPC25 | 0.534 | 73 | C12orf48 | 0.431 |
| 8 | CCNA2 | 0.652 | 41 | KIF18A | 0.528 | 74 | TACC3 | 0.429 |
| 9 | TTK | 0.649 | 42 | PLK1 | 0.508 | 75 | ASF1B | 0.427 |
| 10 | CEP55 | 0.638 | 43 | HMMR | 0.506 | 76 | ERCC6L | 0.424 |
| 11 | CCNB1 | 0.637 | 44 | TOP2A | 0.505 | 77 | TK1 | 0.422 |
| 12 | CDC20 | 0.623 | 45 | CENPF | 0.504 | 78 | TROAP | 0.421 |
| 13 | NCAPH | 0.622 | 46 | ZWINT | 0.502 | 79 | RFC4 | 0.420 |
| 14 | BUB1B | 0.613 | 47 | RAD51AP1 | 0.502 | 80 | PLK4 | 0.420 |
| 15 | KIF23 | 0.602 | 48 | CENPE | 0.498 | 81 | MCM6 | 0.416 |
| 16 | KIF11 | 0.592 | 49 | E2F8 | 0.496 | 82 | KIAA0101 | 0.415 |
| 17 | BIRC5 | 0.588 | 50 | MKI67 | 0.494 | 83 | GINS1 | 0.408 |
| 18 | AURKB | 0.588 | 51 | CENPN | 0.491 | 84 | BLM | 0.406 |
| 19 | NUSAP1 | 0.586 | 52 | CHEK1 | 0.488 | 85 | CKS2 | 0.402 |
| 20 | TPX2 | 0.584 | 53 | MAD2L1 | 0.487 | 86 | MCM2 | 0.402 |
| 21 | RACGAP1 | 0.583 | 54 | GTSE1 | 0.477 | 87 | CENPI | 0.396 |
| 22 | PRC1 | 0.582 | 55 | DTL | 0.475 | 88 | NCAPG2 | 0.393 |
| 23 | ASPM | 0.581 | 56 | RAD51 | 0.474 | 89 | STMN1 | 0.392 |
| 24 | MCM10 | 0.578 | 57 | SHCBP1 | 0.468 | 90 | NEIL3 | 0.389 |
| 25 | NEK2 | 0.573 | 58 | TRIP13 | 0.468 | 91 | ARHGAP11A | 0.387 |
| 26 | UBE2C | 0.569 | 59 | FBXO5 | 0.466 | 92 | ATAD2 | 0.384 |
| 27 | FOXM1 | 0.563 | 60 | FANCI | 0.464 | 93 | CDC6 | 0.383 |
| 28 | CDCA3 | 0.560 | 61 | FEN1 | 0.463 | 94 | CDC7 | 0.378 |
| 29 | NDC80 | 0.555 | 62 | CDC25C | 0.459 | 95 | CCDC99 | 0.377 |
| 30 | STIL | 0.553 | 63 | ECT2 | 0.459 | 96 | CKS1B | 0.375 |
| 31 | KIF15 | 0.553 | 64 | RAD54L | 0.458 | 97 | RNASEH2A | 0.374 |
| 32 | KIF14 | 0.542 | 65 | PBK | 0.456 | 98 | PSRC1 | 0.373 |
| 33 | OIP5 | 0.540 | 66 | KPNA2 | 0.453 | 99 | DONSON | 0.371 |
|  |  |  |  |  |  | 100 | CDC25A | 0.360 |

**Table 3**: Top 100 genes of the lymphocyte-specific attractor based on six datasets

| Rank | Gene Symbol | Avg MI | Rank | Gene Symbol | Avg MI | Rank | Gene Symbol | Avg MI |
|---|---|---|---|---|---|---|---|---|
| 1 | PTPRC | 0.782 | 34 | GPR65 | 0.581 | 67 | NCKAP1L | 0.482 |
| 2 | CD53 | 0.768 | 35 | CD52 | 0.580 | 68 | CD247 | 0.481 |
| 3 | LCP2 | 0.739 | 36 | GIMAP6 | 0.579 | 69 | GZMK | 0.480 |
| 4 | LAPTM5 | 0.707 | 37 | SLAMF8 | 0.578 | 70 | SELL | 0.479 |
| 5 | DOCK2 | 0.699 | 38 | WIPF1 | 0.576 | 71 | LY86 | 0.479 |
| 6 | IL10RA | 0.699 | 39 | MS4A4A | 0.574 | 72 | CCR2 | 0.479 |
| 7 | CYBB | 0.698 | 40 | ARHGAP15 | 0.573 | 73 | ITGAM | 0.478 |
| 8 | CD48 | 0.691 | 41 | CLEC4A | 0.566 | 74 | CORO1A | 0.477 |
| 9 | ITGB2 | 0.679 | 42 | CCL5 | 0.557 | 75 | LILRB1 | 0.473 |
| 10 | EVI2B | 0.675 | 43 | LST1 | 0.557 | 76 | CD74 | 0.473 |
| 11 | MS4A6A | 0.673 | 44 | CD3D | 0.544 | 77 | GPR171 | 0.471 |
| 12 | TFEC | 0.660 | 45 | FGL2 | 0.539 | 78 | HLA-DMB | 0.469 |
| 13 | SLA | 0.657 | 46 | FCGR2B | 0.532 | 79 | ARHGAP25 | 0.468 |
| 14 | SAMSN1 | 0.652 | 47 | MYO1F | 0.530 | 80 | NCF4 | 0.468 |
| 15 | PLEK | 0.649 | 48 | CD163 | 0.524 | 81 | CSF2RB | 0.467 |
| 16 | GIMAP4 | 0.647 | 49 | CLEC7A | 0.521 | 82 | IL7R | 0.464 |
| 17 | GMFG | 0.647 | 50 | CCR1 | 0.517 | 83 | BTK | 0.463 |
| 18 | EVI2A | 0.638 | 51 | HLA-DPA1 | 0.516 | 84 | CD69 | 0.463 |
| 19 | SRGN | 0.637 | 52 | NCF2 | 0.516 | 85 | HLA-DPB1 | 0.455 |
| 20 | AIF1 | 0.636 | 53 | RNASE6 | 0.515 | 86 | ITK | 0.454 |
| 21 | LAIR1 | 0.627 | 54 | CD14 | 0.515 | 87 | TLR1 | 0.454 |
| 22 | FYB | 0.625 | 55 | CD4 | 0.510 | 88 | HLA-DMA | 0.451 |
| 23 | FCER1G | 0.623 | 56 | CD84 | 0.505 | 89 | LILRB4 | 0.449 |
| 24 | CD86 | 0.621 | 57 | NKG7 | 0.504 | 90 | IL2RG | 0.447 |
| 25 | C3AR1 | 0.611 | 58 | C1QA | 0.503 | 91 | TLR2 | 0.446 |
| 26 | C1QB | 0.609 | 59 | TRAF3IP3 | 0.494 | 92 | ALOX5AP | 0.444 |
| 27 | CD2 | 0.606 | 60 | TYROBP | 0.492 | 93 | IFI30 | 0.443 |
| 28 | HCLS1 | 0.599 | 61 | LPXN | 0.492 | 94 | IRF8 | 0.443 |
| 29 | HCK | 0.592 | 62 | IL2RB | 0.490 | 95 | ITGAL | 0.441 |
| 30 | MNDA | 0.588 | 63 | IGSF6 | 0.488 | 96 | TLR8 | 0.438 |
| 31 | CD37 | 0.587 | 64 | CD300A | 0.488 | 97 | PVRIG | 0.437 |
| 32 | CCR5 | 0.585 | 65 | SELPLG | 0.488 | 98 | FLI1 | 0.436 |
| 33 | LY96 | 0.584 | 66 | FCGR2A | 0.486 | 99 | HLA-DRA | 0.434 |
| | | | | | | 100 | LCK | 0.433 |

**Table 4**: List of top ten genes in the chr8q24.3 and HER2 amplicons

| chr8q24.3 | | HER2 | |
|---|---|---|---|
| **Gene** | **Avg MI** | **Gene** | **Avg MI** |
| *PUF60* | 0.694 | *ERBB2* | 0.729 |
| *EXOSC4* | 0.655 | *STARD3* | 0.710 |
| *SHARPIN* | 0.635 | *GRB7* | 0.641 |
| *GPR172A* | 0.620 | *PGAP3* | 0.636 |
| *CYC1* | 0.594 | *MIEN1* | 0.615 |
| *HSF1* | 0.571 | *PSMD3* | 0.494 |
| *FBXL6* | 0.567 | *ORMDL3* | 0.437 |
| *PYCRL* | 0.550 | *GSDMB* | 0.403 |
| *GPAA1* | 0.529 | *MED24* | 0.354 |
| *SCRIB* | 0.483 | *PNMT* | 0.310 |

**Table 5**: Top 50 genes of the estrogen receptor breast cancer attractor

| Rank | Gene Symbol | Avg MI | Rank | Gene Symbol | Avg MI |
|---|---|---|---|---|---|
| 1 | CA12 | 0.596 | 26 | SLC44A4 | 0.344 |
| 2 | AGR2 | 0.567 | 27 | SLC7A8 | 0.339 |
| 3 | GATA3 | 0.553 | 28 | BCL11A | 0.338 |
| 4 | FOXA1 | 0.551 | 29 | FBP1 | 0.338 |
| 5 | MLPH | 0.521 | 30 | MAGED2 | 0.336 |
| 6 | ESR1 | 0.519 | 31 | SLC22A5 | 0.335 |
| 7 | TBC1D9 | 0.502 | 32 | GREB1 | 0.333 |
| 8 | ANXA9 | 0.443 | 33 | EN1 | 0.333 |
| 9 | DNALI1 | 0.424 | 34 | PSAT1 | 0.330 |
| 10 | SCUBE2 | 0.423 | 35 | FOXC1 | 0.329 |
| 11 | NAT1 | 0.422 | 36 | C6orf211 | 0.325 |
| 12 | XBP1 | 0.421 | 37 | C6orf97 | 0.324 |
| 13 | TFF3 | 0.407 | 38 | VGLL1 | 0.321 |
| 14 | ABAT | 0.405 | 39 | CLSTN2 | 0.316 |
| 15 | GFRA1 | 0.397 | 40 | IL6ST | 0.309 |
| 16 | DNAJC12 | 0.395 | 41 | CELSR1 | 0.304 |
| 17 | TFF1 | 0.392 | 42 | TSPAN13 | 0.302 |
| 18 | SLC39A6 | 0.391 | 43 | EVL | 0.302 |
| 19 | SPDEF | 0.382 | 44 | ACADSB | 0.300 |
| 20 | ERBB4 | 0.378 | 45 | SIDT1 | 0.298 |
| 21 | THSD4 | 0.374 | 46 | PGR | 0.296 |
| 22 | MAPT | 0.354 | 47 | C9orf116 | 0.296 |
| 23 | AR | 0.349 | 48 | C14orf45 | 0.288 |
| 24 | DACH1 | 0.348 | 49 | INPP4B | 0.287 |
| 25 | MYB | 0.348 | 50 | AFF3 | 0.287 |

**ONLINE METHODS**

**General attractor finding algorithm**

We chose the association measure $J(G_i, G_j)$ between genes to be a power function with exponent *a* of a normalized estimated information theoretic measure of the mutual information[44] $I(G_i, G_j)$ with minimum value 0 and maximum value 1 (see "Mutual information estimation" below; more sophisticated related association measures[45] can also be used, but computational complexity will be prohibitive). In other words, $J(G_i, G_j) = I^a(G_i, G_j)$, in which the exponent *a* can be any nonnegative number. Each iteration defines a new metagene in which the weight $w_i$ for gene $G_i$ is equal to $w_i = J(G_i, M)$, where *M* is the immediately preceding metagene. The process is repeated until the magnitude of the difference between two consecutive weight vectors is less than a threshold, which we chose to be equal to $10^{-7}$.

At one extreme, if *a* is sufficiently large then each of the seeds will create its own single-gene attractor because all other genes will always have near-zero weights. In that case, the total number of attractors will be equal to the number of genes. At the other extreme, if *a* is zero then all weights will remain equal to each other representing the average of all genes, so there will only be one attractor. The higher the value of *a*, the "sharper" (more focused on its top gene) each attractor will be and the higher the total number of attractors will be. As the value of *a* is gradually decreased, the attractor from a particular seed will transform itself, occasionally in a discontinuous manner, thus providing insight into potential related biological mechanisms.

We empirically found that an appropriate choice of *a* (in the sense of revealing single biomolecular events of co-expressed genes) for general attractors is around 5, in which case there will typically be approximately 50 to 150 resulting attractors, each resulting from many attractee genes. An alternative to the power function can be a sigmoid function with varying steepness, but we found that the consistency of the resulting attractors was worse in that case.

An attractor metagene can also be interpreted as a set of the top genes of the attractor that includes only the genes that are significantly associated with the attractor. One empirical choice includes the genes whose mutual information (or the z-score thereof) with the attractor metagene exceeds a threshold. In fact, the attractor finding algorithm itself can be designed to discover attractor gene sets without assigning weights to genes. In that case, metagenes are defined as simple averages of the genes in a set, and each iteration leads to a new gene set consisting of the new set of top-ranked genes in terms of their association with the previous metagene (gene set sizes can be constant or adaptively changing). This method, however, has the disadvantage of occasionally leading to multiple overlapping attractors.

Identified attractors can be ranked in various ways. The "strength of an attractor" can be defined as the mutual information between the $n^{th}$ top gene of the attractor and the attractor metagene. Indeed, if this measure is high, this implies that at least the top *n* genes of the attractor are strongly co-expressed. We selected *n* = 50 as a reasonable choice, not too large, but sufficiently so to represent a real complex biological phenomenon of co-expression of at least 50 genes. For amplicons, *n* = 5 is sufficient to ensure that the oncogenes are included in the co-expression). We use these choices when referring to the strength of an attractor.

The top genes of many among the found attractors are genomically localized. In that case the biomolecular event that they represent is often the presence of a particular copy number variation. In the cancer datasets that we tried, this phenomenon almost always corresponds to a local amplification event known as an amplicon. We therefore also devised a related amplicon-finding algorithm, custom-designed to identify localized amplicon-representing attractor metagenes, described below.

**Genomically localized attractor finding algorithm**

To identify genomically localized attractors – almost always amplicons – we use the same algorithm but for each seed gene we restrict the set of candidate attractor genes to only include those in the local genomic neighbourhood of the gene, and we optimize the exponent a so that the strength of the attractor is maximized. Specifically, we sort the genes in each chromosome in terms of their genomic location and we only consider the genes within a window of size 51, i.e., with 25 genes on each side of the seed gene. We further optimize the choice of the exponent $a$ for each seed, by allowing $a$ to range from 1.0 to 6.0 with step size of 0.5 and selecting the attractor with the highest strength.

Because the set of allowed genes is different for each seed, the attractors will be different from each other, but "neighbouring" attractors will usually be very similar to each other. Therefore, following exhaustive attractor finding from each seed gene in a chromosome, we apply a filtering algorithm to only select the highest-strength attractor in each local genomic region, as follows: For each attractor, we rank all the genes in terms of their mutual information with the corresponding attractor metagene and we define the range of the attractor to be the chromosomal range of its top 15 genes. If there is any other attractor with overlapping range and higher strength, then the former attractor is filtered out. This filtering is done in parallel, so elimination of attractors occurs simultaneously.

**Mutual information estimation**

Assuming that the continuous expression levels of two genes $G_1$ and $G_2$ are governed by a joint probability density $p_{12}$ with corresponding marginals $p_1$ and $p_2$, the mutual information $I(G_1, G_2)$ is defined as the expected value of $\log(p_{12}/p_1 p_2)$. It is a non-negative quantity representing the information that each one of the variables provides about the other. The pairwise mutual information has successfully been used as a general measure of the correlation between two random variables. We compute mutual information with a spline-based estimator[46] using six bins in each dimension. This method divides the observation space into equally spaced bins and blurs the boundaries between the bins with spline basis functions using third-order B-splines. We further normalize the estimated mutual information by dividing by the maximum of the estimated $I(G_1, G_2)$ and $I(G_1, G_2)$, so the maximum possible value of $I(G_1, G_2)$ is 1.

**Pre-processing gene expression datasets**

We used Level 3 data when directly available, and imputed missing values using a k-nearest-neighbour algorithm with k = 10, as implemented in R[47]. We normalized the other datasets on the Affymetrix platform using the RMA algorithm as implemented in the *affy* package in Bioconductor[48]. To avoid biasing attractor convergence with multiple correlated probe sets of the same gene, we summarized the probe set-level expression values into the gene-level expression values by taking the mean of the expression values of probe sets for the same genes. We used the annotations for the probe sets given in the *jetset* package[49].

To investigate the associations between the attractor metagene expression and the tumor stage and grade, we used the following annotated gene expression datasets. For stage association: Breast (GSE3893), TCGA Ovarian, Colon (GSE14333). For grade association: Breast (GSE3494), TCGA Ovarian, Bladder (GSE13507). For Breast GSE3494 we used only the samples profiled by U133A arrays. For Breast GSE3893 we combined two platforms by taking the intersections of the probes in the U133A and the U133Plus 2.0 arrays. For datasets profiled by Affymetrix platforms all the datasets were normalized using the RMA algorithm. For Bladder GSE13507 normalization was done as provided in the GEO.

***P* value evaluation**

*P* values for gene set enrichment were evaluated with the cumulative hypergeometric distribution using the total number of genes in each dataset.